\documentclass[12pt,preprint]{aastex}

\shorttitle{Bulge Formation in clumpy disks}

\shortauthors{Elmegreen, Bournaud, Elmegreen}

\begin{document}

\title{Bulge Formation by the Coalescence of Giant
Clumps in Primordial Disk Galaxies}

\author{Bruce G. Elmegreen}
\affil{IBM Research Division, T.J. Watson Research Center, 1101
Kitchawan Road, Yorktown Heights, NY 10598, USA} \email{bge@us.ibm.com}
\author{Fr\'ed\'eric Bournaud}
\affil{Laboratoire AIM, CEA-Saclay DSM/IRFU/SAP -
CNRS - Universit\'e Paris Diderot, F-91191~Gif-sur-Yvette Cedex, France}
\email{frederic.bournaud@cea.fr}
\author{Debra Meloy Elmegreen}
\affil{Vassar College, Dept. of Physics \& Astronomy, Box 745,
Poughkeepsie, NY 12604}
\email{elmegreen@vassar.edu}

\begin{abstract}
Gas-rich disks in the early universe are highly turbulent and have
giant star-forming clumps. Models suggest the clumps form by
gravitational instabilities, and if they resist disruption by star
formation, then they interact, lose angular momentum, and migrate to
the center to form a bulge.  Here we study the properties of the bulges
formed by this mechanism. They are all thick, slowly rotating, and have
a high Sersic index, like classical bulges. Their rapid formation
should also give them relatively high $\alpha-$element abundances. We
consider fourteen low-resolution models and four high-resolution
models, three of which have supernova feedback. All models have an
active halo, stellar disk, and gaseous disk, three of the models have a
pre-existing bulge and three others have a cuspy dark matter halo. All
show the same basic result except the one with the highest feedback, in
which the clumps are quickly destroyed and the disk thickens too much.
The coalescence of massive disk clumps in the center of a galaxy is
like a major merger in terms of orbital mixing. It differs by leaving a
bulge with no specific dark matter component, unlike the merger of
individual galaxies. Normal supernova feedback has little effect
because the high turbulent speed in the gas produces tightly bound
clumps. A variety of indirect observations support the model, including
clumpy disks with young bulges at high redshift and bulges with
relatively little dark matter.
\end{abstract}


\section{Introduction}
\label{sect:intro}

Numerical simulations have shown that gas-rich primordial galaxy disks
can fragment into massive clumps that interact gravitationally and
migrate to the center where they merge and form a bulge (Noguchi 1999;
Immeli et al. 2004ab; Carollo et al. 2007). In a recent model with an
initially uniform disk composed of gas and stars and with an active
halo (Bournaud, Elmegreen \& Elmegreen 2007a, hereafter BEE07), clump
formation and interactions trigger a starburst and redistribute the
disk into a smooth exponential profile outside the bulge, while clump
ejecta and clump tidal tails create a steeper exponential profile in
the far outer regions. The earliest stages resemble high-redshift chain
galaxies and clump clusters, with kpc-sized, $10^8$~M$_{\odot}$ clumps
(Cowie et al. 1995; van den Bergh et al. 1996; Elmegreen et al. 2004;
Conselice et al. 2004). The required high gas fractions and disturbed
kinematics are observed at high redshifts (Genzel et al. 2006;
F\"orster Schreiber et al. 2006; Weiner, et al. 2006; Daddi et al.
2008; Bournaud et al. 2008). The final stages in the model resemble
modern spirals with a bulge and generally a double-exponential
(type~II) light profile (Freeman 1970; Pohlen et al. 2002).  The bulge
and thick disk can form at the same time. The bulge mass fraction is
realistic for early type galaxies (BEE07), and, if the disk continues
to grow, it is realistic for late types too. In the present paper, we
explore the detailed properties of bulges that form this way. In a
companion paper (Elmegreen, Bournaud, \& Elmegreen 2008), we propose
that nuclear black holes form along with the bulges by the coalescence
of smaller black holes from each clump.

There are two main types of bulges (e.g., Drory \& Fisher 2007; see
review in Kormendy \& Kennicutt 2004). Early-type galaxies tend to have
``classical'' bulges that are red, old, and high in $\alpha/Fe$ ratio
(indicative of rapid formation), and an $n>2.5$ Sersic profile (Zoccali
et al. 2006; Ballero et al. 2007; McWilliam et al. 2007; Lecureur et
al. 2007). Their age, metallicity and abundance ratios (Moorthy \&
Holtzman 2006), luminosity function (Driver et al. 2007) and position
in the fundamental plane (Falc\'on-Barroso et al. 2002; Thomas \&
Davies 2006; Jablonka et al. 2007; MacArthur et al. 2007) form a
continuous sequence with elliptical galaxies. The most massive bulges
and spheroids are of this type (MacArthur et al. 2007), and the Milky
Way bulge may be also (Rich et al. 2007; Minniti \& Zoccali 2007).
Their origin could be related to galaxy-galaxy mergers (Hernquist \&
Barnes 1991) or cosmological gas accretion (Steinmetz \& Muller 1995;
Xu et al. 2007).  Fu et al. (2003) and Zavala et al. (2007) suggest
that classical bulges form from pre-galactic halo clusters or halo
fragments, respectively, which lose their angular momentum to the halo.

In contrast, late-type galaxies (e.g., later than Sbc - Thomas \&
Davies 2006) tend to have ``pseudo-bulges,'' which are somewhat blue,
span a wide range of ages, and have more disk-like properties,
including small-$n$ Sersic profiles (e.g., Debattista et al. 2006;
Carollo et al. 2007). They presumably formed by secular evolution in
disks, such as vertical instabilities in bars (Combes, \& Sanders 1981;
van den Bosch 1998; Avila-Reese et al. 2005; Debattista et al. 2006;
Athanassoula 2007, 2008), and torque-driven accretion (Pfenniger \&
Norman 1990; Zhang 1999; see review in Kormendy \& Kennicutt 2004).
Rapid secular processes in bars might also build bulges that resemble
the classical type in some respects (Athanassoula \& Martinez-Valpuesta
2007).

The Noguchi model of bulge formation by disk clump coalescence does
not fit either bulge type in standard models. Classical bulges
supposedly formed early without significant disks and then accreted
their disks later (see review in Conselice 2007). This idea is
reinforced by the lack of a correlation between the luminosities of
bulges and disks (Balcells et al. 2007). The Noguchi model, although
fast enough to form a classical bulge in 1 Gyr, requires a disk
first to get the material that coalesces. A pseudo-bulge has a disk
first, but the bulge forms by a different mechanism, usually
involving bars, and it forms over a much longer time than in the
Noguchi model.

The Noguchi model fares better in the context of galactic structure
observed at high redshift.  The disk clumps that are supposed to
coalesce to make a bulge are actually observed in galaxies in the
Hubble Space Telescope Ultra Deep Field (HST UDF) and other deep fields
out to $z\sim6$ (Elmegreen et al. 2007a). These clumps are large ($>1$
kpc), massive ($\sim10^8$~M$_{\odot}$), and young ($\sim300$ Myr, or
$\sim10$ internal dynamical times), making them look like scaled-up
star complexes formed by common gravitational instabilities (Elmegreen
\& Elmegreen 2005; Elmegreen, et al. 2007b). They also contribute
strongly to the gravitational forces in the disk. If they resist
disruption from supernova explosions, then they should migrate to the
center and merge in only a few rotations, which is several $\times10^8$
yrs. The dynamical effect should be analogous to that in a major
merger, with a thorough redistribution of orbits and a starburst in the
gas (BEE07). In this sense, internal clump coalescence in the Noguchi
model is like galaxy coalescence in the classical bulge formation
model. Both processes also get less important over time: bulge
formation by galaxy mergers becomes less important because the galaxy
merger rate decreases; bulge formation by clump mergers becomes less
important because the disks become less clumpy. Disk clumps presumably
result from gravitational instabilities, so their mass depends on the
turbulent Jeans mass. As turbulent speeds decrease relative to the
rotation speed, clump masses decrease relative to the galaxy mass; then
their interactions and migrations toward the bulge becomes less severe.
An important difference between the two cases is that in galaxy
mergers, each component has its own dark matter, whereas in clump
mergers, the components are pure baryons.

A strong constraint on the origin of small bulges in massive galaxies
was recently discovered by Weinzirl et al. (2008), who found too high a
fraction of these bulges compared to the predictions of cosmological
simulations. If a bulge is made by a major merger, as in the classical
description, then the merger had to occur at a redshift $z>2$ in order
for the galaxy to have had enough time to accrete its massive disk
around the merger-remnant bulge. However, the fraction of galaxies with
major mergers only at $z>2$ is much smaller than the fraction of
galaxies with small bulges. Thus another origin for small bulges in
massive galaxies is required. Weinzirl et al. suggested minor mergers
or secular processes are involved. The Noguchi model discussed here is
one possible secular process. Considering the prevalence of the right
initial conditions for this model in deep galaxy surveys, i.e., highly
turbulent and clumpy disks, it may be the primary secular process to
form classical-type bulges.

Genzel et al. (2008) propose that bulges and exponential profiles form
rapidly in young disks following dynamical evolution. They base this on
high resolution observations of H$\alpha$ and [NII] in $z=2$ galaxies
that show highly turbulent motions in star-forming regions and inner
disk masses proportional to age.  Elmegreen et al. (2008) similarly
find an age and bulge-mass sequence in the clumpy galaxies of the
Hubble Space Telescope Ultra Deep Field. The most clumpy types of disks
have younger and smaller bulges relative to their star formation clumps
than the smoother spirals. Elmegreen et al. suggested that clump
cluster and chain galaxies form bulges and thick inner disks while they
are still in their highly turbulent stage. When the gas settles, the
clumps get smaller, the disk gets smoother, and bulge formation slows
down to make a normal spiral.

In the following sections, we first review the observations of high
redshift clumpy galaxies and show examples with and without bulges
(Sect. \ref{sect:obs}). Then we present simulations of gas-rich disks
that form massive star complexes and follow the evolution of those
complexes as they move inward by mutual gravitational interactions
(Sect. \ref{sect:num}). The properties of the resulting bulges are
discussed in Section \ref{sect:results}. The effect of supernova
feedback on the evolution and coalescence of giant clumps is studied in
Section~5. The effect of clump migration on dark matter cusps is
discussed in Section \ref{sect:cusp}. Observations that support this
model of bulge formation are reviewed in Section \ref{sect:discu}.
Section~\ref{sect:sum} contains a summary.

\section{Observations of High-Redshift Clumpy Disks}
\label{sect:obs}

Deep HST images with the Advanced Camera for Surveys (ACS) show two
types of high redshift galaxies that are clumpy and without bulges or
exponential disks: clump-clusters, which are oval collections of bright
clumps, discovered by van den Bergh et al. (1996) in the Hubble Deep
Field, and chains, which are linear collections of clumps, discovered
by Cowie et al. (1995) in HST images of the Hawaii Survey Fields. These
two types have similar luminosities and colors, similar clump
luminosities and colors, and they collectively have a distribution with
respect to their ratio of axes that is constant, as would be the case
for circular disks viewed in random orientations (Elmegreen, et al.
2004; Elmegreen \& Elmegreen 2005). Thus we suggested that most clump
clusters and chains are members of the same population of
highly-clumped disk galaxies, viewed at different orientations
(Elmegreen, et al. 2005a; see also Reshetnikov, Dettmar, \& Combes
2003). Not all clump clusters are single galaxy disks. A few are
composed of features with discrepant redshifts so they are probably
line-of-sight alignments, and a few others look more like separate
galaxies in the process of assembly (see Fig. 5 in Elmegreen et al.
2007b, and Conselice et al. 2008). This is why we prefer to use the
general morphological nomenclature of clump-clusters, rather than
clumpy galaxies. van den Bergh (2002) called them protospirals, and
Conselice et al. (2004) included clump clusters among his more general
class of ``luminous diffuse objects.'' For 178 clump clusters in the
UDF, $27\pm14$\% of the $i_{775}$ light is in clumps, compared to $\sim
8$\% for spirals (Elmegreen et al. 2005b). For ten extreme cases, 40\%
of the $i_{775}$-band light and 19\% of the mass is in clumps
(Elmegreen \& Elmegreen 2005).

The ACS morphology of clumpy galaxies is based on their far-uv
restframe properties, which are dominated by star formation. Rest-frame
images in B-band or redder are required to see bulges. In Elmegreen et
al. (2007a), we compared $i_{775}$ and NICMOS J images to show weak or
no bulges in three clump cluster galaxies, UDF~1666 at redshift
$z=1.318$, UDF~3483 at $z = 1.80$ (photometric), and UDF~6462 at $z =
1.57$, and in one chain galaxy, UDF~7269 at $z=0.69$ (photometric).
Similar comparisons were made in Elmegreen et al. (2008) where the
relative bulge masses were also studied. In this latter paper,
$\sim50$\% of clump clusters and 30\% of chain galaxies in the UDF,
taken from a sample of galaxies larger than 0.3 arcsec, were observed
to have bulges in the H-band NICMOS images.  This leaves more than half
of the clumpy, high-redshift galaxies without bulges.

\begin{figure}
\centering
\includegraphics[width=6.5in]{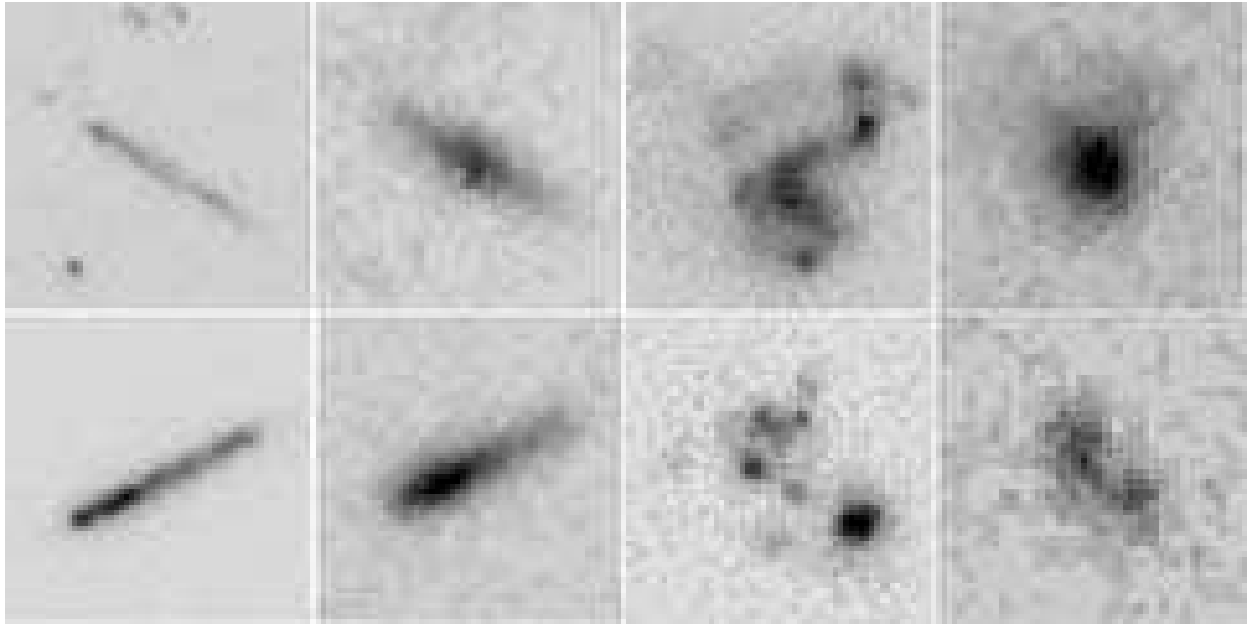}
\caption{Two clumpy galaxies with strong bulges (top) and two with weak
or no bulges (bottom). For each image pair, ACS V-band images are on
the left and NICMOS H-band images (at $3\times$worse resolution) are on
the right. Top left: UDF~6922 ($z_{phot}=1.35$), top right: UDF~4253
($z_{phot}=1.04$), lower left: UDF~7269 ($z_{phot}=0.69$), lower right:
UDF~6486 ($z_{phot}=2.64$). Galaxies in the top row have central bright
concentrations in H-band that are not evident in the V-band, which
shows primarily off-center clumps. Galaxies in the bottom row do not
have bright central concentrations in NICMOS. Four other examples are
in Elmegreen et al. (2007a) and a systematic study of bulgeless clumpy
disks is in Elmegreen et al. (2008). [Image degraded for
astro-ph]}\label{fig:examples}\end{figure}

Different examples of clumpy galaxies with strong and weak bulges are
shown in Figure \ref{fig:examples}.  Each row consists of two galaxies
with ACS V-band images on the left and NICMOS H-band images on the
right. The top left is UDF~6922 ($z_{phot}=1.35$), the top right is
UDF~4253 ($z_{phot}=1.04$), the lower left is UDF~7269
($z_{phot}=0.69$), and the lower right is UDF~6486 ($z_{phot}=2.64$).
The two top galaxies have central bright concentrations in H-band that
are not evident in the V-band, which shows primarily off-center clumps.
These two presumably have bulges.  The two in the bottom row have no
dominant central concentrations in NICMOS. Other images of clumpy
galaxies with and without bulges are shown in Elmegreen et al. (2008).

The identification of bulges in NICMOS images is difficult because the
angular resolution is 3 times worse than in ACS images. The best we can
say until higher resolution IR observations are available is that among
292 clump cluster and chain galaxies with no evident bulges in the ACS
image of the UDF, about half can be examined with NICMOS (because they
are large enough or in the NICMOS field of view) and among these, about
half contain no obvious central bulges even in NICMOS. These galaxies
span redshifts up to $\sim5$ in the UDF and are therefore plausible
examples of the type of pre-bulge clumpy disks discussed in this paper.

It seems reasonable to assume that some high-redshift clumpy galaxies
are fragmented, gas-rich disks. The clumps probably formed by gaseous
gravitational instabilities, as in local galaxies, in which case the
turbulent speed should be moderately large compared to the rotation
speed, perhaps several tens of percent. This makes the clump size large
compared to the galaxy radius, so there are only a few giant clumps in
each galaxy. The high turbulence is in agreement with observations by
Genzel et al. (2006), F\"orster Schreiber et al.(2006), and others
referenced above. One of the clump-clusters that has a small bulge or
reddish stellar region, UDF~6462, has a global rotation curve
indicative of a single galaxy with highly disturbed velocities at the
positions of the clumps, suggesting both high turbulence and
non-circular motions (Bournaud et al. 2008). The bulge in UDF~6462 is
redder than the clumps, somewhat centralized, and has a higher
metallicity than the rest of the disk.

\section{Numerical simulations} \label{sect:num}

The most relevant observations for a model of clumpy galaxies are the
overall galaxy size, and the clump masses, separations and numbers. If
the clumps form by gravitational instabilities in the gaseous component
of an irregular disk, then their properties depend on local values of
the turbulent speed $\sigma$ and column density $\Sigma$. To get a
small number of massive clumps, we need a relatively high ratio of
turbulent speed to rotation speed, $\sim10$\% or more, and to get such
a turbulent disk gravitationally unstable so that it forms clumps, we
need a fairly high gas column density, which means a high gas-to-star
ratio in the disk at that time. Many other details of the initial
conditions are less important for this bulge formation model. The
clumps should have a mass comparable to the disk Jeans mass, which is
$M_{J}\sim \sigma_{gas}^4/\left(G^2\Sigma\right)$, and they should have
a separation comparable to the Jeans length,
$L_J\sim2\sigma_{gas}^2/\left(G\Sigma\right)$. If the dispersion for
gas is large, comparable to that for the disk stars, then $\Sigma$ in
this equation will have a strong contribution from disk stars, i.e.,
both stars and gas can contribute to the clump. When the Jeans length
is a large fraction of the disk scale length, the clump masses will be
a large fraction of the disk mass, and such clumps will interact
strongly (BEE07). Cosmological simulations have more realism than our
models on extragalactic scales, but they are not generally tuned to
give clumpy disks like those we see at intermediate to high redshifts.
Analytical models that approximate disk accretion with viscosity and
consider thermal cooling (e.g., Lodato \& Natarajan 2006) are also not
as useful as direct simulations for following the complexity of
turbulent disk fragmentation and evolution.

The evolution of gas-rich galaxy disks is modeled here in the same
manner as in BEE07. We use a particle-mesh sticky-particle code (see
Bournaud \& Combes 2002, 2003). Sticky-particle parameters are
$\beta_r=\beta_t=0.7$ for all but run 4, which has
$\beta_r=\beta_t=0.8$. Star formation follows a local Schmidt law where
the probability per timestep that each gas particle is transformed into
a stellar particle is proportional to the local gas density to the
power 1.4 (e.g., Kennicutt 1998). The proportionality factor gives a
star formation rate of $3.5$~M$_{\sun}$~yr$^{-1}$ in the initially
uniform disk.

In all our models, the initial disk is composed of gas and stars with a
uniform surface density (i.e., chain and clump cluster galaxies do not
have exponential disks; Elmegreen et al. 2005b). The initial disk
radius is 6~kpc and the thickness is $h=700$~pc with a
$\mathrm{sech}^2\left( z/h \right)$ vertical distribution. The disk
mass (stars+gas) is $7 \times 10^{10}$~M$_{\sun}$. Stars have a Toomre
parameter in the stable regime, $Q_{\mathrm{s}}=1.5$, but the mixed
disk of stars and gas is unstable.

We first analyze a series of 14 low-resolution runs with a grid
resolution and gravitational softening length of 110~pc. Stars, gas,
and dark matter halo are modeled with one million particles each in
these runs. Feedback from star formation is not included. Runs~0 to 7
are the same as in BEE07, and their initial parameters were summarized
in Table 1 of BEE07. The main fixed parameters were recalled above. The
gas mass fraction in the disk, $f_{\mathrm{G}}$, is 0.5 except for runs
4 and 5, where it is 0.25 and 0.75. The halo-to-disk mass ratio, $H/D$,
inside the initial disk radius, is 0.5 except for runs 6 and 7, where
it is 0.25 and 0.80.

Runs~0 to 7 have no initial bulge and a dark halo that is a Plummer
sphere with a scale-length of 15~kpc. In addition, runs 0N, 1N, 2N have
a $\Lambda$CDM cuspy halo (Navarro, Frenk \& White 1997) with a cusp
scale-length $r_S = 6$~kpc (concentration parameter 16.7 for a virial
radius of 100~kpc). The other parameters including the gas fractions
and halo-to-disk mass ratios are unchanged from runs 0, 1 and 2,
respectively. Runs 0B, 1B, 2B are like runs 0, 1, and 2, but they have
a small initial bulge that is a Plummer sphere with 10\% of the disk
mass and a radial scale-length of 600~pc.

We also performed higher-resolution models, with some including
supernova feedback. The number of particles of each type was larger by
a factor 3, the grid resolution and softening length were improved by a
factor 2 in each direction everywhere, and they were improved by a
factor 4 at radii smaller than 4~kpc. Clump coalescence occurs in this
inner region, so the high resolution cases, where the spatial
resolution is 28~pc, consider this process best.

With high resolution, Run~HR-0 has an initial gas fraction of 60\%, an
initial bulge of 4\% of the disk, and the other parameters as in the
fiducial Run~0. Runs~HR-0-F1, HR-0-F2 and HR-0-F3 have the same initial
conditions and resolution as run HR-0, but with supernova feedback
included. We used the feedback model proposed by Mihos \& Hernquist
(1994). A supernova is assumed to occur for each $50\;M_{\odot}$ of
stars formed. A fraction $\epsilon$ of the $10^{51}$~erg energy of each
supernova is released in the form of radial velocity kicks applied to
gas particles within the closest cells. We used $\epsilon =
2\times10^{-4}$ in run HR-0-F1 (which is about the value suggested by
Mihos \& Hernquist 1994), $\epsilon = 10^{-3}$ in run HR-0-F2 and
$\epsilon = 10^{-2}$ in run HR-0-F3.

During the course of each simulation, clumps were identified every
25~Myr as regions where the surface density is locally larger than the
radial average by a factor of 3. Only clumps with masses larger than
$2\times10^8$ $M_\odot$ and sizes smaller than 3~kpc were considered:
this size limitation avoids spiral arms (see BEE07). Sub-structures
like OB associations and small dense clusters are expected to form
inside the main clumps of real galaxies, but we cannot resolve this in
the simulations.

\section{Bulge Properties} \label{sect:results}

Properties of the bulges formed in the low and high resolution runs
without feedback are first considered. The influence of supernovae
(SNe) in runs HR-0-F1 to HR-0-F3 will be discussed in section
\ref{sect:sn}. The time-dependent mass column density in the fiducial
run~0 was shown in Figure 4 of our previous paper, BEE07. Giant stellar
clumps formed in the gas-rich initial disk and then migrated to the
center where they made a bulge. About half the clump mass was left in
the disk and the other half got to the center. An exponential disk
formed outside.

\begin{figure}
\centering
\includegraphics[width=6in]{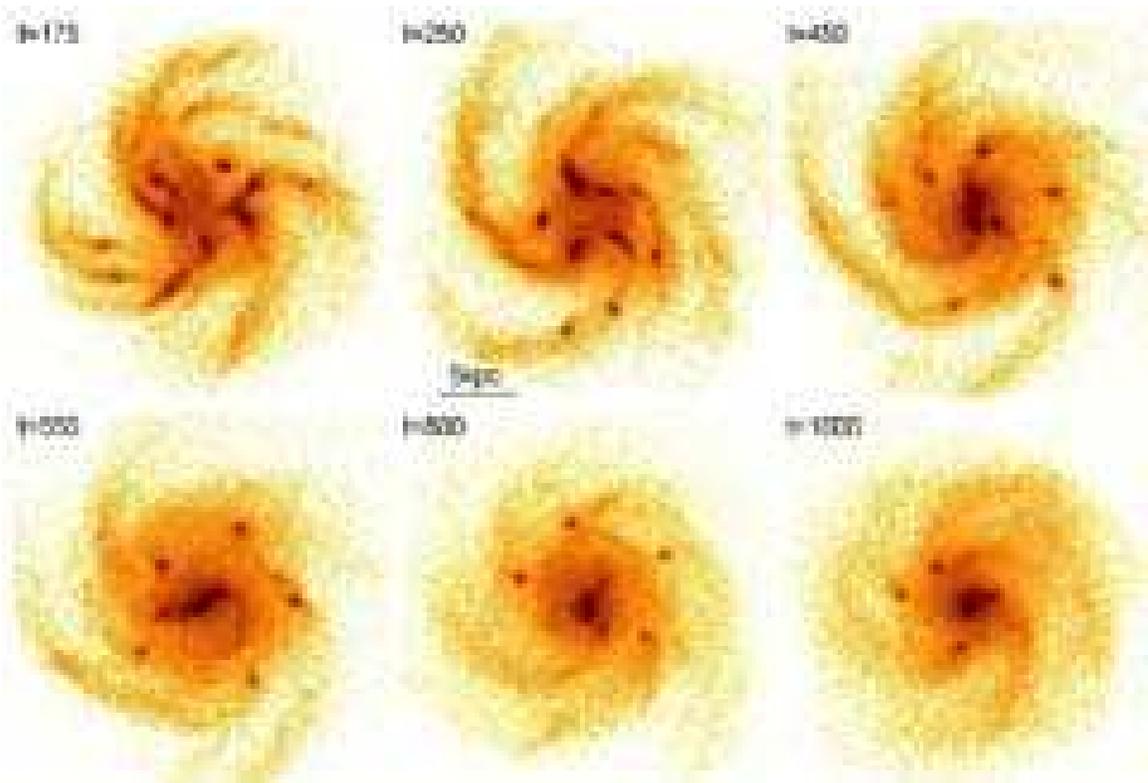}
\caption{Face-on snapshots of the disk mass density (gas and stars) for
run~0N, which has a cuspy dark matter profile. Time is in Myr. Clumps
form quickly in the disk and move to the center, where they coalesce
into a bulge within 1 Gyr. Extra star formation in the bulge region is
triggered at the time of merging too. A few clumps remain in the disk
when the simulation ends. [Image degraded for
astro-ph]}\label{fig:run0N}\end{figure}

\begin{figure}
\centering
\includegraphics[width=6in]{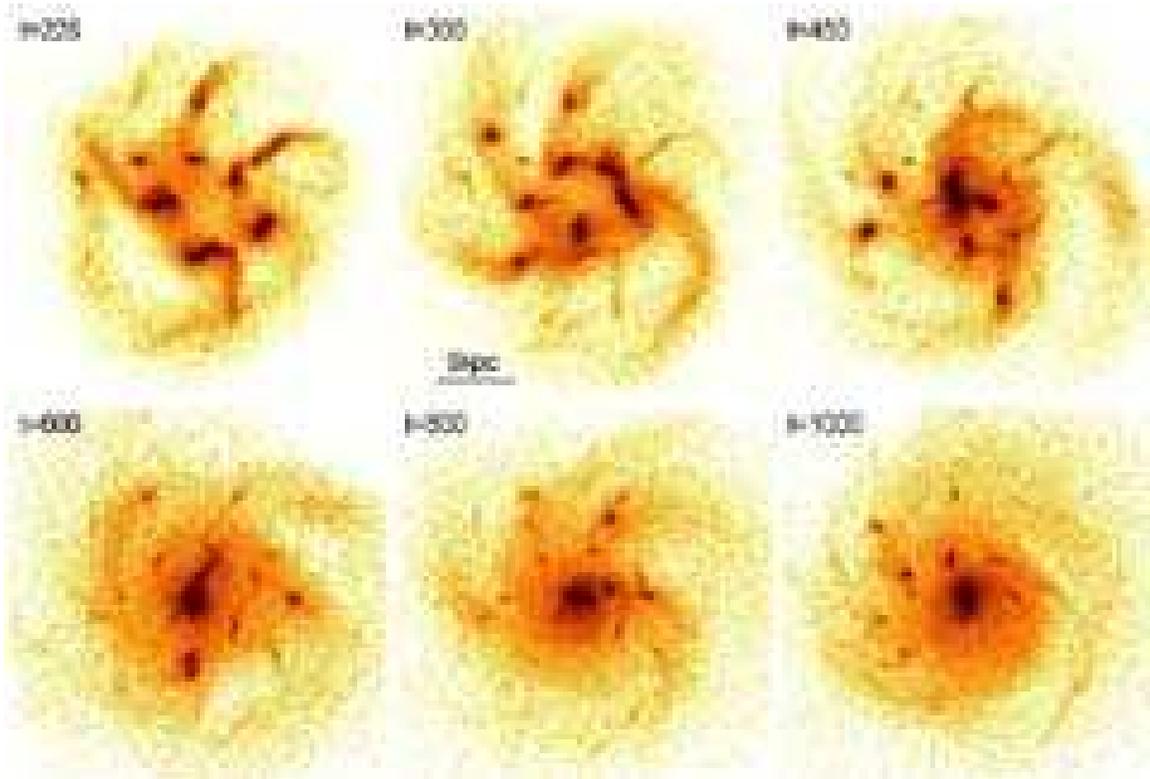}
\caption{Same as Fig. 2 but for run~0B, where a primordial bulge of
10\% of the disk mass is initially present in the model. The initial
bulge stars are not shown here; only the gas and stars from the initial
disk are shown. [Image degraded for
astro-ph]}\label{fig:run0B}\end{figure}

Figures \ref{fig:run0N} and \ref{fig:run0B} show the new fiducial runs,
0N, with an initial NFW halo, and 0B, with an initial central bulge.
The timescale for clump merging has increased slightly in both cases
compared to run~0 because of the relatively smaller mutual forces
between the clumps: smaller compared to the dark matter force in run~0N
and smaller compared to the combined disk+bulge force in run~0B. The
evolution of the high resolution run HR-0 is shown on the top in
Figures~\ref{fig:hr1} and \ref{fig:hr2}.

\subsection{Bulge mass} \label{sect:bulgemass}

The fractional clump masses, $f_C$, bulge masses, and bulge-to-total
(baryonic) mass fractions after 1 Gyr are given in Table 1. Runs 0 to 7
and HR-0 have bulge fractions ranging from 0.12 to 0.36 (average 0.24).
Runs with a cuspy dark matter halo have about the same final bulge mass
fractions as those with a Plummer sphere halo. Runs with an initial
10\% bulge have slightly larger final bulge masses than those without
an initial bulge. This increase is not the full 10\% because increased
tidal forces near the center disrupt the clumps earlier before they
coalesce. Bulge growth is somewhat regulated in this way. This keeps
the final bulge-to-disk ratio compatible with that in present-day
early-type spirals, even when a primordial bulge is present from an
earlier clump-cluster phase or another formation process.

\subsection{Bulge Density Profile}

\begin{figure}
\centering
\includegraphics[width=4in]{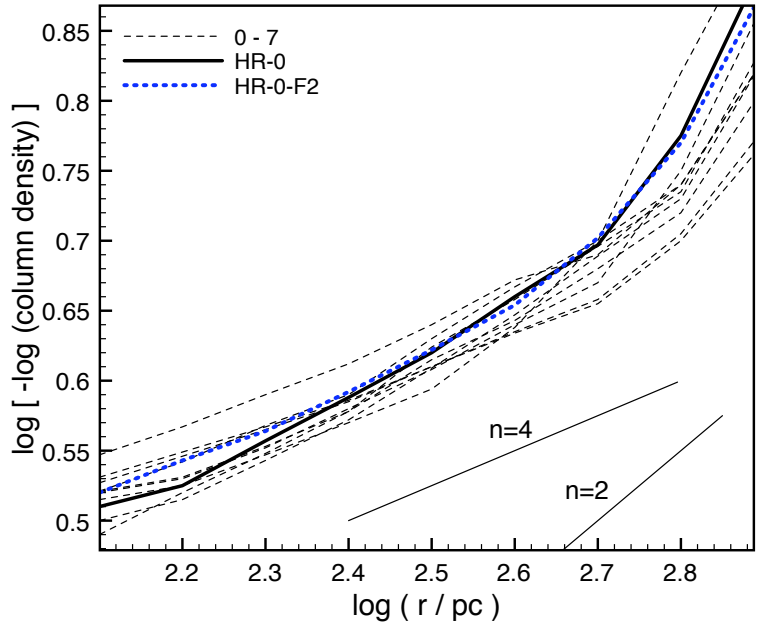}
\caption{Normalized column density profile in the inner 800~pc plotted
so that the slope is inversely proportional to the Sersic index, $n$.
The bulge profiles have a high Sersic index at radii of 100-500~pc,
indicating they resemble classical bulges. The thin dashed lines are
runs 0 to 7, the thick solid line is run HR-0 (higher resolution) and
the thick dotted line is run HR-0-F2 (supernova feedback).
}\label{fig:profile}\end{figure}

Figure \ref{fig:profile} shows radial disk column density profiles,
($\Sigma(r)$, normalized to the central value) of the bulge regions up
to $\sim 1.5$~kpc, where the disks become exponential. We plot
$\log[-\log(\Sigma)]$ versus $\log(r)$ because that has a constant
slope of $1/n$ in a Sersic profile with projected normalized column
density $\Sigma=\exp[-(r/r_0)^{1/n}]$. The model bulges have steep
Sersic indices, with an average of $n\simeq 4$ (between 3 and 4.5) in
the 100--500~pc radial range. The Sersic index decreases at radii of
500 to 1000~pc. The bulge in the higher-resolution run HR-0 also has a
Sersic index clearly larger than 2 and close to 4 for $r<500$~pc; the
best fit in the 100-500~pc range is $n=3.4$. Thus the model produces
central concentrations with the morphology of classical bulges.

\subsection{Bulge Kinematics}

\begin{figure}
\centering
\includegraphics[width=6in]{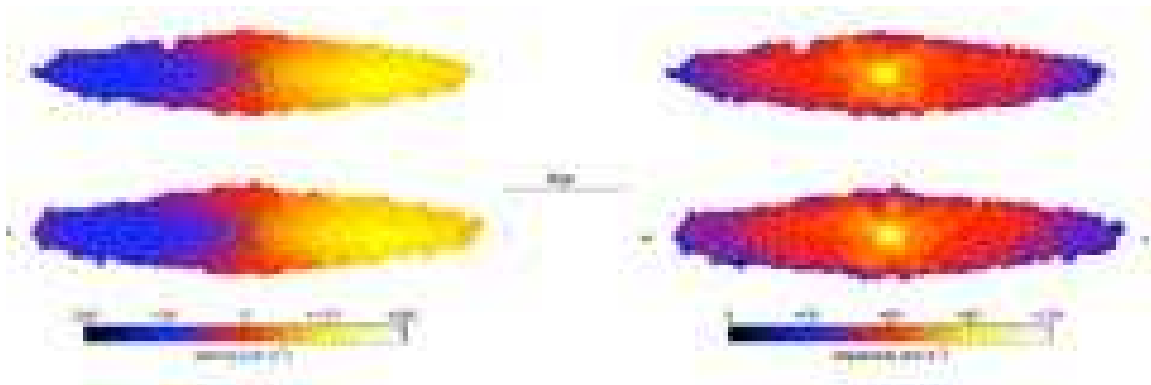}
\caption{Line-of-sight velocity (left) and velocity dispersion(right)
maps for two orthogonal projections of run~0 at the final stage. The
green circles have a radius of 750~pc, which contains about two thirds
of the bulge mass. The central regions have high velocity dispersions
and slow rotations, indicating that clump coalescence is like a major
merger with a randomization of orbits and a loss of angular momentum to
the disk. [Image degraded for astro-ph] }\label{fig:Bkin}\end{figure}

The bulges formed by clump coalescence are not fast rotators. Figure
\ref{fig:Bkin} shows the line-of-sight velocity ($V$) and velocity
dispersion ($\sigma_{bulge}$) for the fiducial run~0 at the final
stage, considering two orthogonal edge-on projections. The bulge
central velocity dispersion was measured as follows: for each model, we
built 50 independent 2-D projections of the stellar content, uniformly
distributed over the sine of the inclination to represent an average
over all possible viewing angles. The line-of-sight velocity dispersion
was computed inside a projected aperture diameter of 110~pc (the grid
resolution), and for each model we kept the average value over all
projections. We do not attempt to subtract the disk component from
edge-on projections, nor to make a direct 3-D measurement, because this
would not be done in observations. Measuring the mass-weighted
line-of-sight dispersion within a larger projected aperture of 500~pc,
we find values between 7\% higher and 11\% lower than the measurement
in the central 110~pc. Thus the projected radial gradient in the
dispersion is relatively small in the bulge. This result was also found
observationally by Gebhardt et al. (2000), who showed that the
dispersion gets high only in the very central regions close to a black
hole.

Figure \ref{fig:Bkin} indicates that the bulge region has a strong
increase in the velocity dispersion compared to the disk. While the
$V/\sigma_{bulge}$ ratio in the exponential disk exceeds 2, the
luminosity-weighted average value in the central 750~pc, which contains
two-thirds of the bulge mass (green circles on Fig.~5), is much
smaller, between 0.4 and 0.5 for the two projections. This is actually
an upper limit to $V/\sigma_{bulge}$ in the bulge because of
unavoidable disk contamination. In this respect, clump mergers that
form bulges are like galaxy mergers that form elliptical galaxies. Both
provoke a large kinematical heating that forms a spheroid dominated by
random motions. Clump mergers are not like the in-spiral of small dense
clusters from dynamical friction, which can leave a rapidly rotating
core. Values of the bulge line-of-sight velocity dispersion for all
simulations are in Table 1. The average is 130 km s$^{-1}$.

The bulge velocity dispersion was not affected by resolution. Even at
higher resolution, the bulges are slowly rotating and have high
velocity dispersions. The bulge morphology and kinematics in model HR-0
are shown in Figure~\ref{fig:hr3}. This bulge is rotating with $V
\leq$~60~km~s$^{-1}$, and it has a velocity dispersion $\sigma
\geq$~80~km~s$^{-1}$ within its half-mass radius.

All of the bulges are somewhat flattened in both the high and low
resolution runs. This flattening results from a combination of slight
bulge rotation, anisotropic velocity dispersion, and a flattened disk
potential.

\section{Effects of Supernova Feedback on the Evolution of High-Redshift Clumpy
Galaxies}\label{sect:sn}

\begin{figure}
\centering
\includegraphics[width=6in]{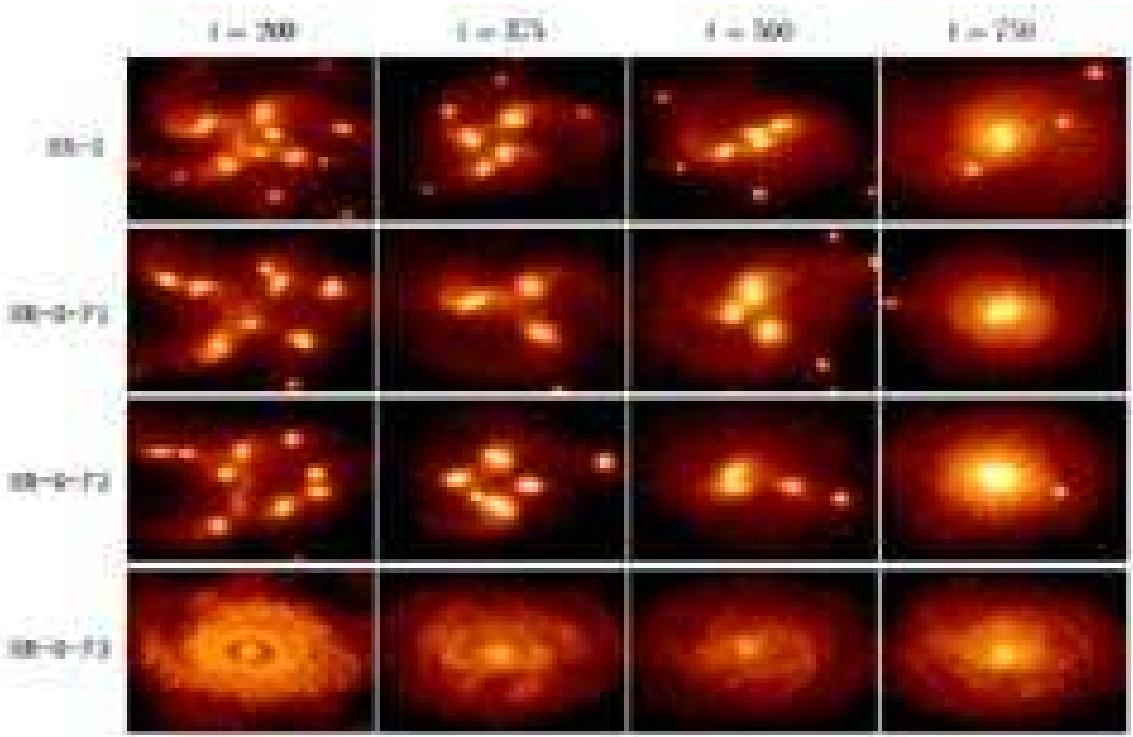}
\caption{High-resolution models without (top) and with supernova
feedback. The feedback efficiency increases from the second row to the
bottom row, with such a high efficiency in the run HR-0-F3 that all
star-forming clumps are rapidly destroyed. The snapshots show 30-degree
projections of the total mass density, in log scale, and are $14 \times
8$~kpc wide. [Image degraded for astro-ph]}\label{fig:hr1}\end{figure}

\begin{figure}
\centering
\includegraphics[width=6in]{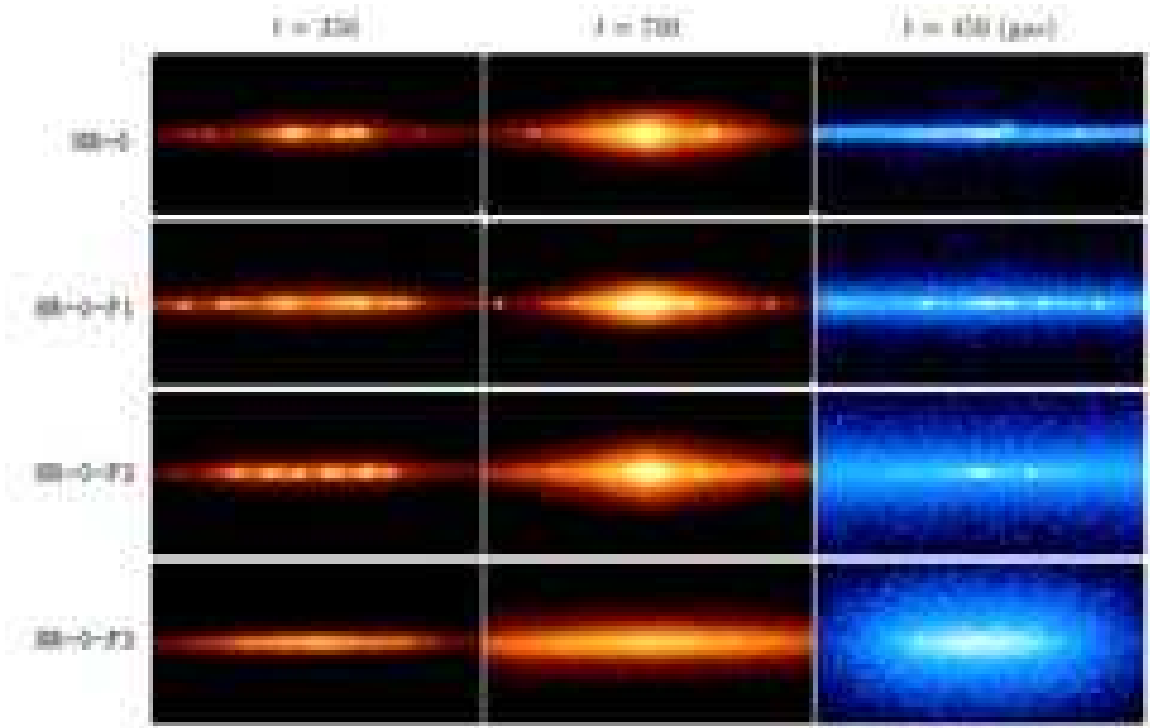}
\caption{High-resolution models without (top) and with supernova
feedback in edge-on projections, for the stellar mass density at
$t=350$ and $700$~Myr, and for the gas mass density at $t=450$~Myr. The
snapshots are in log scale, and are $15 \times 7$~kpc wide. [Image
degraded for astro-ph]}\label{fig:hr2}\end{figure}

\begin{figure}
\centering
\includegraphics[width=4in]{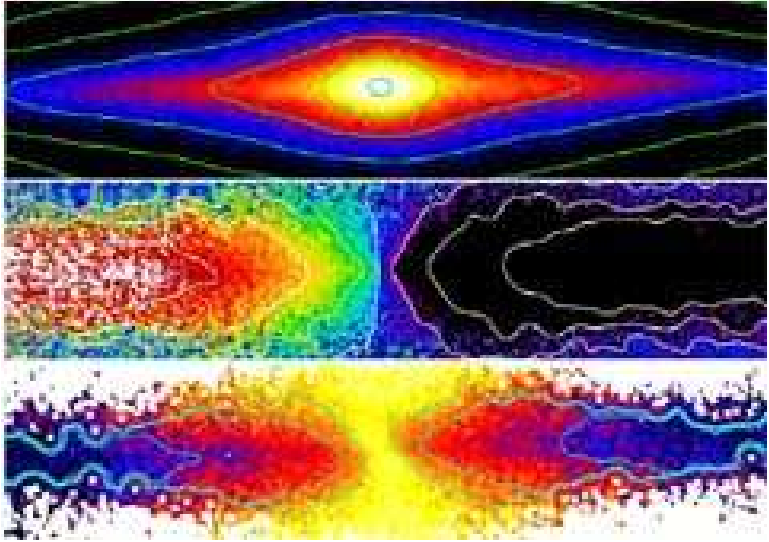}
\caption{Edge-on view of the bulge in the high-resolution run HR-0 at
the end of the simulation ($t=1.3$~Gyr). The snapshots show the mass
density (top), velocity field (middle) and line-of-sight velocity
dispersion (bottom). Contours on the velocity field are at $V=0$, $\pm
35$, $\pm 70$ and $\pm 105$~km~s$^{-1}$; $\sigma=40$, $70$, and
$100$~km~s$^{-1}$ on the dispersion map. Comparable results from a low
resolution run were shown in Fig. 5. [Image degraded for
astro-ph]}\label{fig:hr3}\end{figure}

The models described so far and in BEE07 include star formation but no
SN feedback. The concern is whether energy injected by SNe can disrupt
the clumps before they reach the center. Immeli et al. (2004a,b) had
feedback in their models and the clumps were not disrupted, however
their clump masses were larger than in the observations and in our
models. Here we consider three high-resolution models, HR-0-F1,
HR-0-F2, and HR-0-F3, that have different fractions $\epsilon$ of SN
energy re-injected into the local gas kinetic energy
($\epsilon=2\times10^{-4}$, $10^{-3}$, and $10^{-2}$, respectively).
The evolution of these models is shown in the bottom three rows of
Figures~\ref{fig:hr1} and \ref{fig:hr2}.

According to Mihos \& Hernquist (1994), realistic values for $\epsilon$
in kinetic feedback models should be around $2\times 10^{-4}$ (our run
HR-0-F1) and not higher than $10^{-3}$ (our run HR-0-F2). These two
runs have long-lived, massive clumps of gas and stars in their disks
that survive until they coalesce into the central bulge
(Figs.~\ref{fig:hr1} and \ref{fig:hr2}). Only the lowest mass clumps
($10^7$~M$_{\sun}$) are significantly reduced by feedback. As measured
in Table~1, the masses of the largest clumps are only slightly reduced
when feedback is added, and as a result the bulge growth is reduced
too, both by about 10\%. This is a rather small effect, so feedback is
not the main driver of clump disruption. As shown in BEE07, clumps
typically release half of their mass into the disk before they reach
the bulge because of tidal forces and shear. This keeps the final
bulge-to-disk mass ratio rather low. In our models HR-0-F1 and HR-0-F2,
SN feedback further regulates the bulge growth in a modest way, but the
cores of the clumps clearly survive until they reach the central kpc
and coalesce into the bulge.

The edge-on view of the gas disk in run HR-0-F2 ($\epsilon = 10^{-3}$;
Fig. ~\ref{fig:hr2}) shows that it is significantly thickened. At the
mid-course of the simulation, the gas layer has an average scale-height
of 1.6~kpc, larger than the initial conditions and thicker than the
star-forming layers in chain galaxies (Elmegreen \& Elmegreen 2006).
This model should then be taken as an upper limit on the feedback
efficiency, in agreement with what Mihos \& Hernquist inferred at lower
redshift. Even in this case, though, the cores of the massive clumps
survive until they reach the central bulge. If we increase even more
the feedback efficiency (run HR-0-F3), the clumps become short-lived:
they form in the gas but are rapidly disrupted, with no visible
counterpart remaining in the stellar mass. In this case no bulge
formation by clump coalescence can occur. The gas disk is also nearly
destroyed in this case:  the thickening is so large that the gas disk
resembles a spheroid and even the stellar disk thickens over time.

It is thus unlikely that supernova feedback can disrupt $10^8\;M_\odot$
clumps in high-redshift galaxies; otherwise the gas disk itself would
get dispersed. Assuming a lower feedback efficiency but a higher star
formation efficiency, or a higher supernova fraction in the IMF, would
lead to the same conclusion. If the energy input from stellar feedback
is low enough to preserve the gas disk, then the most massive clumps
that form are likely to survive for a few disk rotations and reach the
center. Supernovae feedback should be more important in lower-mass
galaxies, which should have lower gas turbulent speeds and easier clump
disruption. This could explain why lower mass galaxies (i.e., later
Hubble types) tend to have lower bulge-to-disk mass ratios.

\section{Evolution of the Dark Matter Cusp} \label{sect:cusp}

\begin{figure}
\centering
\includegraphics[width=6in]{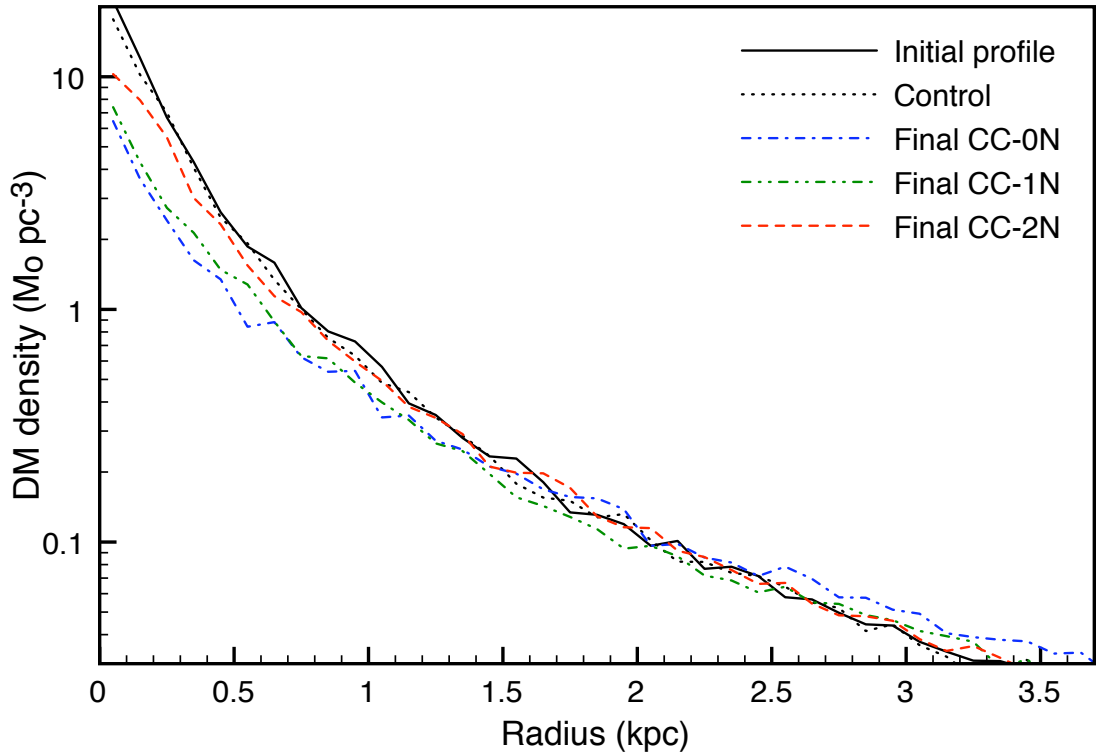}
\caption{Density profiles of the dark matter in the three
simulations started with NFW haloes (solid line). The three final
profiles are at 1~Gyr, when the clumps have finished there
migration, disruption and merging. The control run is with a rigid
disk potential, showing that the profile evolution on the other runs
is not a numerical artifact from the relaxation of the model halo,
but caused by the clump-cluster evolution. The final profiles have a
lower central peak density and somewhat larger scale-length, but are
still cuspy.}\label{fig:DMprof}\end{figure}

Three of the models have a cuspy initial dark matter halo, following
Navarro et al. (1997). The evolution of the disk does not depend much
on this cusp, as shown above, but the evolution of the cusp depends a
little on the disk. Figure \ref{fig:DMprof} shows the halo density
profiles before and after these three simulations. A control run with a
rigid disk potential is also shown (dotted line). When the disk can
evolve through the formation and interaction of clumps, the central
cusp becomes considerably weaker, amounting to a factor of $\sim3$ in
central halo density. The cusp scale-length increases also. This change
is a result of halo heating by the moving massive clumps when they
reach the center. The central cusps do not go away, but they get
reduced. Their reduction should be greater in lower mass galaxies if
the disk Jeans mass is about the same (i.e., if the gas velocity
dispersion and column density are about the same for a smaller mass
galaxy). The reduction should be greater also if several episodes of
gas accretion, clump formation and bulge growth take place during the
life of a galaxy. Each episode would erode the cusp a little more.

\section{Discussion}\label{sect:discu}

Numerical simulations suggest that primitive gas-rich disks should
fragment into clumpy star-formation complexes with masses of
$\sim10^8\;M_\odot$ or more, and that these complexes should interact
gravitationally and move to the galaxy center where they combine to
form a bulge. The surrounding disk becomes exponential in the process.
There are many assumptions here, the simulations are highly idealized,
and the initial galaxies are not fully cosmological with realistic gas
accretion and hierarchical build-up. Still, the most important aspects
of the model are rather fundamental: young galaxies are gas-rich,
gravitational instabilities are inevitable (as in modern galaxies), and
clump interactions follow from the relatively large Jeans masses. Our
models also suggest that the most massive clumps can survive
star-formation feedback during the time it takes them to coalesce.
Thus, we are led to ask whether this bulge formation scenario is
compatible with the properties of real galaxies and if there is either
direct or indirect evidence for it.

First, we have illustrated the kinds of young galaxies where these
processes should operate: the clump cluster and chain galaxies with
weak or no bulges that are seen out to the detection limit at $z \sim
5$ in all high-resolution deep images (Sect. \ref{sect:intro},
\ref{sect:obs}). The observed clump masses, sizes, positions, numbers,
and ages in these galaxies are reproduced by the models without fine
tuning. The key ingredient is a ratio of velocity dispersion to
rotation speed that is several times larger than in local galaxies:
this large dispersion makes the interstellar Jeans mass equal to the
observed clump mass for star formation that is triggered by
gravitational instabilities (e.g., Elmegreen et al. 2007b). Such large
dispersions are observed in high redshift disks (Sect. 1).

Sargent et al. (2007) also note that the co-moving density of
bulge-free disks among intermediate-size galaxies increases out to
$z=1$, while the co-moving density of classical bulgy disks (as
determined from their high-$n$ Sersic profiles) decreases, keeping the
total disk comoving density about the same. They conclude that at least
some high-$n$ bulges have formed by secular disk processes since $z=1$.
Balcells \& Dom\'inguez-Palmero (2007) find a similar coeval evolution
of classical bulges and disks: bulge colors correlate with disk colors
and red bulges are rarely surrounded by purely blue disks in the
redshift range from 0.1 to 1.2. Thus there is good evidence for
bulge-free disks at high redshift, and for a transition from bulge-free
to classical-bulge galaxies over time. This is compatible with the
standard model of bulge formation in galaxy mergers, and also with our
model of bulge formation by baryonic clump mergers.

The simulations show that the bulges formed this way have properties
typical of classical, early-type galaxy bulges. They are slow rotators
with high velocity dispersions, thick perpendicular dimensions, and
high Sersic indexes. The model timescale from the formation of giant
clumps to their merging in the bulge is about 1~Gyr, consistent with
the evidence summarized in Section 1 for rapid formation of classical
bulges, including that in the Milky Way (e.g., Lecureur et al. 2007)
and $z\sim2$ galaxies (Genzel et al. 2008; Elmegreen et al. 2008).
Longer bulge formation timescales are possible if the clump formation
phase in the disk is prolonged, as might be the case with continued
cosmological gas accretion.

The merger of disk clumps is somewhat similar to the merger of small
galaxies. The initial angular momentum of the clumps is lost to the
disk and halo, so all that remains after their merger is a spheroidal
bulge supported mainly by velocity dispersions. An important difference
between bulge formation by clump mergers and bulge formation by galaxy
mergers concerns the dark matter content of the merging entities. Disk
clumps, forming by disk instabilities, contain no halo dark matter from
the surrounding galaxy, and so the bulges they form contain no
additional dark matter either. In contrast, galaxy mergers bring in
considerable amounts of dark matter with them, and much of this should
end up in the bulge or spheroid (e.g. Dekel et al. 2005). The amount of
bulge dark matter depends on how many bulge stars form at the time of
the merger: merger-driven gas accretion followed by star formation to
make a bulge brings in relatively little dark matter, whereas merging
warm stellar populations should preserve their initial dark matter
content. Because observations of early-type galaxy bulges show
relatively little dark matter (Corradi \& Capaccioli 1990, Noordermeer
et al. 2007, Corsini 2007), the galaxy merger model of bulge formation
could have more difficulty than the clump merger model in this respect.

Another test of our model involves dark matter cusps (Navarro et al.
1997). Dark matter in merging galaxies adds to the cusp during
hierarchical buildup, while massive baryonic clumps that stir the
central regions reduce these cusps by increasing the central dark
matter velocity dispersion (see also El-Zant, Shlosman, \& Hoffman
2001; Sect. \ref{sect:cusp}). Neither of these two bulge-forming
mechanisms can completely remove the cusps. Either real galaxies never
had cusps, or their cusps have been washed out by various dynamical
mechanisms. The merging of disk clumps into a bulge eases the cusp
problem, while the merging of cusp-containing galaxies makes the
problem worse.

The case for classical bulge formation by galaxy coalescence has been
weakened also by the observation that early-type bulges may have
smaller Sersic $n$ than previously thought. Carollo, Stiavelli, \& Mack
(1998) found nuclear star clusters in half of the 75 galaxies they
observed with HST, spanning a wide range of Hubble types, and noted
that many had exponential profiles outside the clusters and few had
pure $n=4$ Sersic profiles. Balcells et al. (2003) also found for
early-type galaxies (S0-Sbc) that nuclear star clusters, which are
usually unresolved from the ground, convert low-$n$ profiles into
high-$n$ profiles artificially. High-$n$ profiles can be explained by
major mergers and violent relaxation (van Albada 1982; Barnes 1988;
Bournaud et al. 2005, Naab \& Trujillo 2006) or by numerous minor
mergers (Bournaud et al. 2007b), but if the profiles have lower $n$,
then the merger scenario is less compelling. Our models form high-$n$
bulges by clump mergers and are subject to the same criticism. If all
bulges have low $n$ except for unresolved nuclear clusters, then our
model cannot explain them -- or strong supernova feedback is required
to decrease the mass concentration. More high-resolution bulge profiles
are required for this important test.

In the case of late-type or low-mass bulges, which are mostly
pseudo-bulges, there is considerable evidence that they come from
disks, including correlations between bulge colors and scale lengths
and disk colors and scale lengths (MacArthur et al. 2003; Carollo et
al. 2001; 2007; see Kormendy \& Kennicutt 2004). The implied processes
usually involve torque-driven inflows and vertical resonances from bars
and spirals. Our models differ from this in that they are fast and
require a more primitive disk, one dominated by turbulent gas and
little pre-existing bulge, bar, or spiral potentials. A predominantly
stellar disk with a bar already, or with enough of a stellar mass to
generate strong spiral torques, is not as likely as a primitive disk to
fragment into giant clumps that interact gravitationally and merge in
the center. We showed in Section \ref{sect:bulgemass} that an initial
bulge increases the tidal force on the clumps and limits the mass they
can deliver to the center.

\section{Conclusions}
\label{sect:sum}

Gas-rich disks with relatively high turbulent speeds should fragment
into massive, star-forming clumps and quickly evolve to resemble the
chain and clump-cluster galaxies observed at high redshift. Further
evolution should lead to clump coalescence and the overall structure of
early-type disk galaxies, with smooth (Type II) exponential profiles
and Sersic $n\sim4$ bulges. The bulge and inner thick disk stars form
fast enough to become enriched with $\alpha$ elements, as required, and
their velocity dispersions are realistically large. Direct observations
in support of this model include clumpy, highly-turbulent disks and the
presence of relatively young, low-mass bulges at intermediate to high
redshifts (Sects. 1, 2). Indirect observations include a gradual
transition in the classical bulge fraction in disks from redshift
$z\sim1$ to 0 (Sect. \ref{sect:discu}), a high fraction of massive
galaxies with relatively low-mass bulges (Sect. 1), and a lack of dark
matter specific to the bulge component (Sect. \ref{sect:discu}), all of
which constrain bulge formation by major mergers of individual
galaxies. We also note some tendency for clump interactions to heat a
dark matter cusp if there is one, and in our simulations, this lowers
the cusp density by a factor of $\sim3$. More simulations will be
required to determine if multiple bulge-building episodes continue to
reduce the cusp density.

The primary two motivations to consider bulge formation by this
mechanism are the large turbulent speeds and the extremely clumpy
structures observed in high redshift galaxies.  Gravitational
instabilities in highly turbulent disks will form stars in extremely
massive clumps, so the two observations go together. The present
simulations start at this point and show that such clumps will interact
strongly with the disk stars and with each other and, as a result, lose
angular momentum and migrate to the center. Bulge formation follows.
Our simulations with supernova feedback indicate that the massive
clumps survive long enough for this whole process to occur, even if
feedback regulates bulge growth slightly. The models also indicate that
clump coalescence is a lot like galaxy coalescence in that the final
bulges formed by these two mechanisms have similar velocity
dispersions, rotations, light profiles, and three-dimensional
structures. Bulge formation by clump coalescence is expected to get
less frequent over time as disk turbulence subsides, because the
star-forming clumps will get less massive with the lower turbulent
speeds.  This decrease in bulge formation by clump coalescence
parallels the decrease expected for bulge formation by galaxy
coalescence.

The dynamical processes shown here are fundamental yet rapid,
suggesting they may be present but unrecognized in other published
simulations, including large-scale cosmological simulations that have
enough resolution to see structure in disk galaxies (e.g., Rasera \&
Teyssier 2006; Ocvirk et al. 2008). If the clumpy-disk, bulge-formation
phase lasts for only a few disk rotations, and most of the time in a
simulation consists of calm secular phases alternating with minor and
major mergers, then this important first step in the formation of a
galaxy might be missed compared to the subsequent longer-term
evolution. The contribution of such clumpy phases to the star formation
history might be overlooked too. Still, the processes described here
should be present in simulations with enough resolution.

\acknowledgments

Numerical simulations were carried out on the NEC-SX8R vector computer
at CEA/CCRT. D.M.E. thanks Vassar College for publication support.
Helpful comments by the referee are gratefully acknowledged.

{}

\clearpage

\clearpage

\begin{table}
\centering
\caption{Masses, Mass fractions, and Bulge Velocity
Dispersions}
\begin{tabular}{ccccccccccc}
\tableline\tableline
Run ID & $f_C$ & Bulge Mass & Bulge/Total & $\sigma_{bulge}$\\
  &     &  $\times10^9\;M_\odot$  & Mass Ratio & km s$^{-1}$\\
\tableline
0   &0.38  &21    &0.30    &121\\
1   &0.32  &14.7   &0.21    &117\\
2   &0.31  &13.3   &0.19    &108\\
3   &0.39  &23.1  &0.33    &146\\
4   &0.26  &9.8    &0.14     &87\\
5   &0.34  &22.4  &0.32    &184\\
6   &0.41  &25.2  &0.36    &178\\
7   &0.23  &8.4    &0.12     &91\\
0N  &0.33  &19.6  &0.28    &163\\
1N  &0.29  &12.6   &0.18    &118\\
2N  &0.32  &14.7   &0.21    &122\\
0B  &0.31  &24.6 &0.32    &145\\
1B  &0.26  &20.8 &0.27    &125\\
2B  &0.28  &19.3 &0.25    &109\\
HR-0 & 0.36 & 19 & 0.26 & 123\\
HR-0-F1 & 0.33 & 18 & 0.25 & 119\\
HR-0-F2 & 0.30 & 14 & 0.20 & 107\\
\tableline
\end{tabular}
\tablecomments{
1. Run ID. Runs with N start with a NFW halo profile, runs with B start with an initial bulge.\\
2. Mass fraction in clumps at the time of peak clumpiness.\\
3. Bulge mass (measured as in BEE07).\\
4. Final bulge-to-total mass ratio, not including dark matter in the total.\\
5. Bulge central line-of-sight velocity dispersion averaged from 50
projections uniformly distributed over the sine of the inclination
angle.}
\end{table}

\clearpage

\end{document}